\documentclass[conference]{IEEEtran}

\usepackage{graphicx}
\usepackage{cite}
\usepackage{subfigure}
\usepackage{amsmath}
\usepackage{float}
\usepackage{stfloats}
\usepackage{algorithmic}
\usepackage{algorithm}

\IEEEoverridecommandlockouts
\begin{document}

\title{\huge Performance Analysis of Spectrum Handoff for Cognitive Radio Ad Hoc Networks without Common Control Channel under Homogeneous Primary Traffic}
\normalsize
\author{\authorblockN{Yi Song and Jiang Xie}
\authorblockA{Department of Electrical and Computer Engineering \\
The University of North Carolina at Charlotte \\
Email: \{ysong13, jxie1\}@uncc.edu}\thanks{This work was supported in part by the US National Science Foundation (NSF) under Grant No. CNS-0855200, CNS-0915599, and CNS-0953644.}}

\IEEEaftertitletext{\vspace{-2.0\baselineskip}}
\maketitle
\begin{abstract}
Cognitive radio (CR) technology is regarded as a promising solution to the spectrum scarcity problem. Due to the spectrum varying nature of CR networks, unlicensed users are required to perform spectrum handoffs when licensed users reuse the spectrum. In this paper, we study the performance of the spectrum handoff process in a CR ad hoc network under homogeneous primary traffic. We propose a novel three dimensional discrete-time Markov chain to characterize the process of spectrum handoffs and analyze the performance of unlicensed users. Since in real CR networks, a dedicated common control channel is not practical, in our model, we implement a network coordination scheme where no dedicated common control channel is needed. Moreover, in wireless communications, collisions among simultaneous transmissions cannot be immediately detected and the whole collided packets need to be retransmitted, which greatly affects the network performance. With this observation, we also consider the retransmissions of the collided packets in our proposed discrete-time Markov chain. In addition, besides the random channel selection scheme, we study the impact of different channel selection schemes on the performance of the spectrum handoff process. Furthermore, we also consider the spectrum sensing delay in our proposed Markov model and investigate its effect on the network performance. We validate the numerical results obtained from our proposed Markov model against simulation and investigate other parameters of interest in the spectrum handoff scenario. Our proposed analytical model can be applied to various practical network scenarios. It also provides new insights on the process of spectrum handoffs. Currently, no existing analysis has considered the comprehensive aspects of spectrum handoff as what we consider in this paper.
\end{abstract}

\IEEEpeerreviewmaketitle
\section{Introduction}
\label{sc:introduction}
According to the frequency chart from the Federal Communications Commission (FCC), almost all radio spectrum suitable for wireless communications (3kHz-300GHz) has been allocated. However, recent studies indicate that much of the radio spectrum is not in use for a significant amount of time and in a large number of locations. A report from FCC shows that up to 85\% of the assigned spectrum is underutilized due to the current static spectrum allocation policy \cite{FCC-2003}. This spectrum underutilization leads to a new dynamic spectrum allocation paradigm approved by FCC which allows users to exploit temporal and spatial spectrum holes or white spaces in a radio environment \cite{Akyildiz-Lee06}. Cognitive radio (CR) is considered as a key technology to implement dynamic spectrum access (DSA) that allows an unlicensed user (or, secondary user) to adaptively determine appropriate operating parameters to access the licensed spectrum band which is not occupied by licensed users (or, primary users) \cite{Mitola00}.

Since secondary users (SUs) are regarded as visitors to the licensed spectrum, cognitive radio networks bring about unique challenges in designing novel spectrum management functions because of their coexistence with legacy networks. One critical challenge is that SUs should avoid causing harmful interference to primary users (PUs) and support seamless communications regardless of the appearance of PUs. In particular, one of the most important functionalities of CR networks is \textit{spectrum mobility} which refers to the capability of SUs to change the operating frequencies based on the availability of the spectrum. Spectrum mobility gives rise to a new type of handoff called \textit{spectrum handoff} which refers to the process that when the current channel used by a SU is reclaimed by a PU, the SU needs to pause the on-going transmission, vacate that channel, and determine a new available channel to continue the transmission. Needless to say, spectrum mobility is essential for the performance of SU communications. However, most existing work on CR networks focus on other functionalities of CR networks (\textit{spectrum sensing}, \textit{spectrum management}, and \textit{spectrum sharing}) \cite{Akyildiz-Lee06}, while spectrum mobility is less investigated in the research community. Various models have been proposed to address the performance analysis issue of the other three functionalities \cite{Fanwang08,Chang08,Jafar2007,Tang07}, but not spectrum mobility.

Related work on spectrum handoffs in CR networks falls into two categories based on the moment when SUs carry out spectrum handoffs. In the first category, SUs perform channel switching \textit{after} detecting the reappearances of PUs, namely the \textit{reactive} approach \cite{Willkomm05,LCWang09,CWWangGC10}. In the other category, SUs predict the future PU channel activities and perform spectrum handoffs \textit{before} the disruptions with PU transmissions, namely the \textit{proactive} approach \cite{Zheng-proactive08,Clancy-2006,Arslan-predict07,Yoon10icc}. With the exception of \cite{LCWang09} and \cite{CWWangGC10}, the performance analysis of all prior works on spectrum handoffs is simulation-based.

An analytical model is of great importance for performance analysis because it can provide useful insights on the operation of spectrum handoffs. However, there have been limited studies on the performance analysis of spectrum handoffs in CR networks using analytical models. In \cite{LCWang09} and \cite{CWWangGC10}, a preemptive resume priority queueing model is proposed to analyze the total service time of SU communications for proactive and reactive-decision spectrum handoffs. However, in both \cite{LCWang09} and \cite{CWWangGC10}, only one pair of SUs is considered in a network, while the interference and interactions among SUs are ignored, which may greatly affect the performance of the network. Additionally, although they are not designed for the spectrum handoff scenario, some recent related works on analyzing the performance of SUs using analytical models can be found in \cite{SWanginfocom10} and \cite{Laoinfocom10}. In \cite{SWanginfocom10}, a dynamic model for CR networks based on stochastic fluid queue analysis is proposed to analyze the steady-state queue length of SUs. In \cite{Laoinfocom10}, the stationary queue tail distribution of a single SU is analyzed using a large deviation approach. In all the above proposals, a common and severe limitation is that the detection of PUs is assumed to be perfect (i.e., a SU transmitting pair can immediately perform channel switching if a PU is detected to appear on the current channel, thus the overlapping of SU and PU transmissions is negligible). However, since the power of a transmitted signal is much higher than the power of the received signal in wireless medium due to path loss, instantaneous collision detection is not possible for wireless communications. Thus, even if only a portion of a packet is collided with another transmission, the whole packet is wasted and need to be retransmitted. Without considering the retransmission, the performance conclusion may be inaccurate. Unfortunately, it is not easy to simply add retransmissions in the existing models. In this paper, we model the retransmissions of the collided packets in our proposed Markov model. To the best of our knowledge, this is the first paper that considers the retransmissions of the collided packets in spectrum handoff scenarios.

Furthermore, in the prior proposals, the network coordination and rendezvous issue (i.e., before transmitting a packet between two nodes, they first find a common channel and establish a link) is either not considered\cite{LCWang09}\cite{CWWangGC10}\cite{Clancy-2006}\cite{Arslan-predict07}\cite{SWanginfocom10}\cite{Laoinfocom10} or simplified by using a dedicated common control channel (CCC)\cite{Willkomm05}\cite{Zheng-proactive08}\cite{Yoon10icc}. Since the dedicated CCC is always available, a SU can coordinate with its receiver at any moment when there is a transmission request. However, it is not practical to use a dedicated CCC in CR networks because it is difficult to identify a dedicated common channel for all the SUs throughout the network since the spectrum availability varies with time and location. In this paper, we do not make such assumption. We model the scenario where SUs need to find an available channel for network coordination. Therefore, in this paper, we consider a more practical distributed network coordination scheme in our analytical model design.

In this paper, we also explore the effect of different channel selection schemes on the performance of the spectrum handoff process using our proposed Markov model. Besides the general random channel selection scheme, different channel selection schemes have been proposed for various design goals \cite{LCWang09}\cite{Tang08}\cite{SongGC10}. These channel selection schemes can be easily adopted in our proposed three dimensional discrete-time Markov chain if we apply different state transition probabilities in the proposed analytical model.

In addition, we also consider the impact of the spectrum sensing delay on the performance of the spectrum handoff process. Since the overlapping time of a SU and PU collision is not negligible, we define the spectrum sensing delay as the duration from the moment a collision happens to the time a SU detects the collision. The spectrum sensing delay can be easily employed in our proposed Markov model with minor modifications.

In summary, in this paper, we study the performance of SUs in the spectrum handoff scenario in a CR ad hoc network where PU traffic on each channel is identical. The main contributions of this paper are given as follows:
\begin{enumerate} \item We propose a novel three dimensional Markov model to characterize the process of spectrum handoffs and analyze the performance of SUs. The interference and interactions among multiple SUs are considered in our proposed model.
\item Due to the spectrum-varying nature of CR networks, in our model, we implement a more practical coordination scheme in our proposed model instead of using a dedicated CCC to realize channel rendezvous. 
\item Since instantaneous collision detection is not feasible for wireless communications, we consider the retransmissions of the collided SU packets in spectrum handoff scenarios.
\item We apply three different channel selection schemes in the proposed Markov model and study their effects on the performance of SUs in spectrum handoff scenarios. 
\item We consider the spectrum sensing delay and its impact on the network performance. This feature can be easily implemented in our proposed Markov model. \end{enumerate}
Therefore, our model is very flexible and can be applied to many practical scenarios with various designs.

The rest of this paper is organized as follows. In Section \ref{sc:networkmodel}, the network coordination scheme and the spectrum handoff process considered in this paper are introduced. In Section \ref{sc:analysis}, a three dimensional discrete-time Markov model is proposed. In Section \ref{sc:selection} and \ref{sc:sensing}, the performance analysis for different channel selection schemes and the spectrum sensing delay is given. Numerical results using the proposed Markov model are presented in Section \ref{sc:evaluation}, followed by the conclusions in Section \ref{sc:conclusion}.

\section{Network Coordination and Spectrum Handoff}
\label{sc:networkmodel}
\subsection{Network Coordination Scheme}
\label{ssc:coordination}
\begin{figure}[htb!]
\vspace{-0.1in}
\centerline{\includegraphics[width=0.49\textwidth]{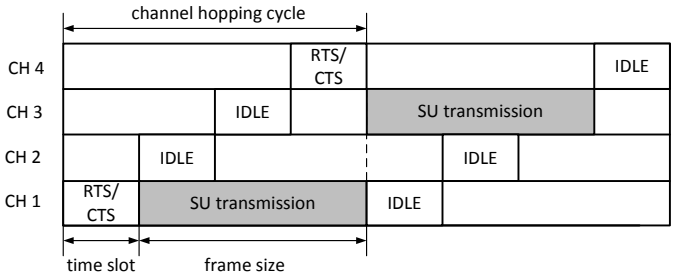}}
\vspace{-0.05in}
\caption{An example of the network coordination scheme.}
\label{fig:coordination}
\vspace{-0.13in}
\end{figure}

Throughout this paper, we consider a network scenario where $N$ pairs of SUs form a CR ad hoc network and opportunistically access $M$ identical licensed channels. We use the common frequency-hopping sequence approach as the network coordination scheme  \cite{Zhang02}\cite{Tang99}. Fig. \ref{fig:coordination} illustrates the operations of the common frequency-hopping sequence approach, where the channels are time-slotted and SUs communicate with each other in a synchronous manner. This is similar to the frequency hopping technique used in Bluetooth. When no packet needs to be transmitted, all SUs are required to follow the same channel-hopping sequence to hop through the band (e.g., the hopping pattern cycles through channels $1,2,\cdots,M$).  If a pair of SUs wants to initiate a transmission, they first exchange request-to-send (RTS) and clear-to-send (CTS) packets during a time slot. Then, after the SU transmitter successfully receives the CTS packet, they pause the channel hopping and start data transmissions on an available channel, while other non-transmitting SUs continue the channel-hopping. The selected channel information is contained in the RTS packet. After the data being successfully transmitted, the SU pair should switch back to the channel-hopping sequence and rejoin the channel-hopping. In this paper, we define the length of a time slot as the transmission delay of sending one RTS/CTS pair. We also assume that spectrum sensing is perfect (i.e., SUs can sense all the channels simultaneously and always make the correct decisions).

In this paper, we assume that any SU data packet is transmitted at the beginning of a time slot and ends at the end of a time slot. This implies that the length of a SU packet is a multiple of a time slot. This assumption is commonly used in time-slotted systems \cite{Su08}\cite{Su-ciss-08}. We further define that a SU packet is segmented into frames and each frame contains $c$ time slots. At the end of a frame, the two SUs can either rejoin the channel hopping when a data transmission ends, or start another frame on the same channel. Therefore, if a SU packet collides with a PU packet, only the collided frame will be retransmitted while the successfully received frames will not be retransmitted. Thus, the probability of successfully transmitting a whole packet is improved.

\subsection{Spectrum Handoff Process}
\label{ssc:spectrumhandoff}
\begin{figure}[htb!]
\vspace{-0.23in}
\centerline{\includegraphics[width=0.36\textwidth]{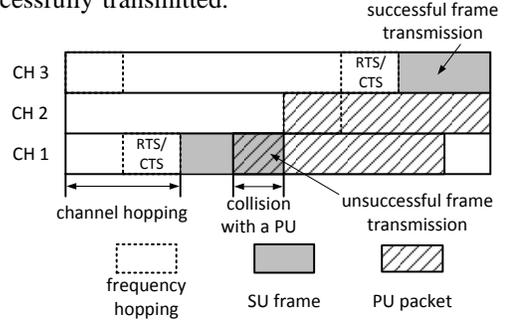}}
\vspace{-0.08in}
\caption{An example of the spectrum handoff process.}
\label{fig:handoff}
\vspace{-0.12in}
\end{figure}

Fig. \ref{fig:handoff} shows an example of a spectrum handoff process considered in this paper in a three-channel scenario. Before a data transmission starts, SUs hop through the channels following the same frequency-hopping sequence. Once a successful RTS/CTS handshake between a SU transmitter and its receiver takes place, the two SUs pause the channel hopping and start the data transmission. If a PU packet transmission starts in the middle of a SU transmission, the transmitter cannot instantaneously detect the collision. Thus, the SU transmitting pair will know the collided transmission till the end of the frame (e.g., the transmitter does not receive the acknowledgment (ACK) from the receiver). Then, the two SUs resume the channel hopping for coordination until they find another idle channel for the retransmission of the previously unsuccessful frame. On the other hand, if a SU frame does not collide with a PU packet, the SU transmitter continues to transmit the next frame on the same channel until all frames are successfully transmitted.

\section{The Proposed Three Dimensional Discrete-time Markov Model}
\label{sc:analysis}
In this section, we develop a Markov model to analyze the performance of the spectrum handoff process. For simplicity, we assume the same number of neighbors per SU, which permits us to focus on any SU to analyze the performance. We ignore the propagation delay or any processing time in our analysis. We also assume that the destination of any data packet from a SU transmitter is always available, that is, the probability that the selected SU receiver is not busy is one.
\subsection{The Proposed Markov Model}
\label{ssc:markov}
Based on the time slotted channels, any action of a SU can only be taken at the beginning of a time slot. In addition, the status of a SU in the current time slot only relies on its immediate past time slot. Such discrete-time characteristics allow us to model the status of a SU using Markov chain analysis. From Fig. \ref{fig:handoff}, the status of a SU in a time slot can only be one of the following:
\begin{enumerate}
	\item \textit{Idle}: no packet arrives at a SU.
	\item \textit{Transmitting}: the transmission of a SU does not collide with PU packets in a time slot, i.e., successful transmission.
	\item \textit{Collided}: the transmission of a SU collides with PU packets in a time slot, i.e., unsuccessful transmission.
	\item \textit{Backlogged}: a SU has a packet to transmit in the buffer but fails to access a channel.
\end{enumerate}
Note that there are two cases that a SU can be in the \textit{Backlogged} status. In the first case, when a SU pair initiates a new transmission, if multiple SU pairs select the same channel for transmissions, a collision among SUs occurs and no SU pair can access the channel. Thus, the packet is backlogged. Similarly, in the second case, when a SU pair performs a spectrum handoff, if multiple SU pairs select the same channel, a collision among SUs occurs and the frame in every SU is also backlogged.

\begin{figure}[htb!]
\vspace{-0.05in}
	\centering
		\includegraphics[width=0.49\textwidth]{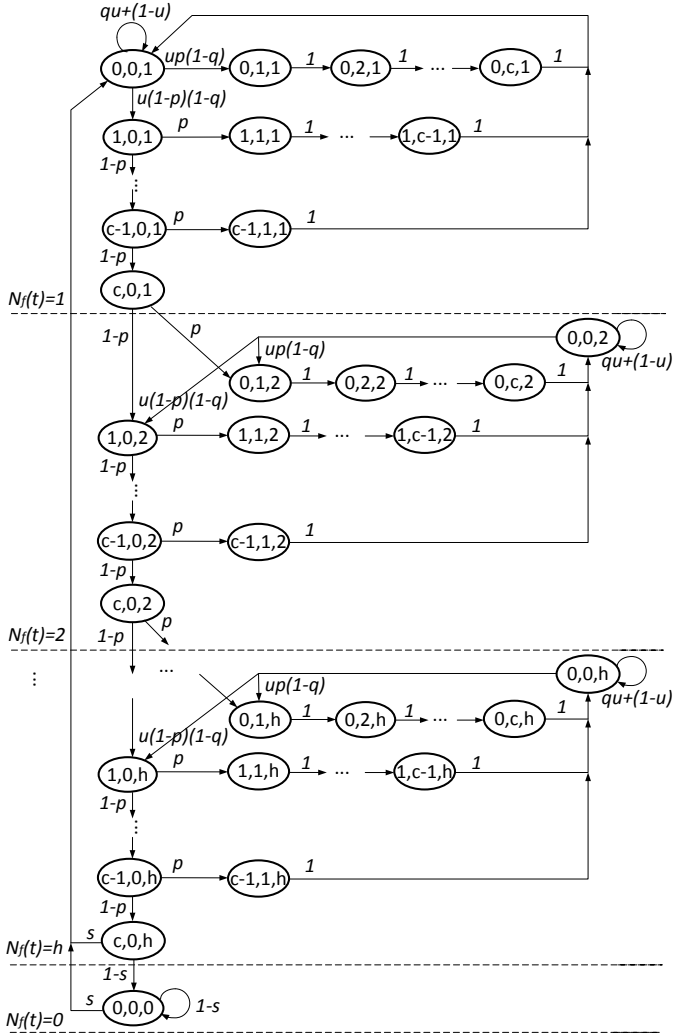}
	\caption{The transition diagram of the proposed Markov model.}
	\label{fig:markov}
	\vspace{-0.09in}
\end{figure}

As mentioned in Section \ref{sc:introduction}, we consider the scenario that when a collision between a SU and PU happens, the overlapping of a SU frame and a PU packet is not negligible. Thus, the number of time slots that a SU frame collides with a PU packet is an important parameter to the performance of SUs. Based on the above analysis, the state of the proposed Markov model at time slot $t$ is defined by a vector $(N_t(t),N_c(t),N_f(t))$, where $N_t(t),N_c(t), {\rm and~} N_f(t)$ denote the number of time slots including the current slot that are successfully transmitted in the current frame, the number of time slots including the current slot that are collided with a PU packet in the current frame, and the number of frames that have been successfully transmitted plus the current frame that is in the middle of a transmission at time slot $t$, respectively. Therefore, $N_t(t)\!+\!N_c(t)\!\leq \!c$. Fig. \ref{fig:markov} shows the state transition diagram of our proposed three dimensional Markov chain. There are totally $(h\!+\!1)$ tiers in the state transition diagram. For each tier, it is a two dimensional Markov chain with a fixed $N_f(t)$. Table \ref{tb:notation} summarizes the notations used in our Markov model.
\begin{table}[htb]\caption{Notations Used in the Markov Analysis}
\centering
\begin{tabular}{|c|l|}
\hline
Symbol & Definition \\
\hline
$p$ & Probability that a PU packet arrives in a time slot \\
\hline
$s$ & Probability that a SU packet arrives in a time slot \\
\hline
$h$ & Number of frames in a SU packet \\
\hline
$c$ & Number of time slots in a frame\\
\hline
$q$ & Probability of a collision among SUs\\
\hline
$u$ & Probability that at least one channel is idle \\
\hline
\end{tabular}
\label{tb:notation}
\vspace{-0.15in}
\end{table}

From Fig. \ref{fig:markov}, it is observed that the proposed Markov model accurately capture the status of a SU in a time slot. The state $(N_t(t)\!\!=\!\!0,N_c(t)\!\!=\!\!0,N_f(t)\!\!=\!\!0)$ in Fig. \ref{fig:markov} represents that a SU is in the $Idle$ status. Similarly, the states $(N_t(t)\!\in\!\![1,c],N_c(t)\!=\!0,N_f(t)\!\in\![1,h])$ represent the $Transmitting$ status, i.e., no collision. The states $(N_t(t)\in[0,c-1],N_c(t)\in[1,c],N_f(t)\in[1,h])$ represent the $Collided$ status. At last, the states $(N_t(t)\!\!=\!\!0,N_c(t)\!\!=\!\!0,N_f(t)\in[1,h])$ represent the $Backlogged$ status, where $(N_t(t)\!\!=\!\!0,N_c(t)\!\!=\!\!0,N_f(t)\!\!=\!\!1)$ is the \textit{Backlogged} status during a new transmission. As shown in Fig. \ref{fig:markov}, the feature of the common frequency-hopping sequence scheme is captured in our model that a SU can only start a new transmission when there is a channel available. In the following discussion, we use the terms ``states'' in our proposed Markov model and the ``status'' of a SU in a time slot interchangeably. We also use the notations $(N_t(t\!+\!1)\!\!=\!\!i,N_c(t\!+\!1)\!\!=\!\!j,N_f(t\!+\!1)\!\!=\!\!k)$ and $(i,j,k)$ to represent a state interchangeably.
\vspace{-0.06in}

\subsection{Derivation of Steady-State Probabilities}
\label{ssc:prob}
To obtain the steady-state probabilities of the states in the three dimensional Markov chain shown in Fig. \ref{fig:markov}, we first get the one-step state transition probability. We denote the one-step state transition probability from time slot $t$ to $t+1$ as $P(i_1,j_1,k_1|i_0,j_0,k_0)\!\!=\!\!P(N_t(t\!+\!1)\!\!=\!\!i_1,N_c(t\!+\!1)\!\!=\!\!j_1,N_f(t\!+\!1)\!\!=\!\!k_1|N_t(t)\!\!=\!\!i_0,N_c(t)\!\!=\!\!j_0,N_f(t)\!\!=\!\!k_0)$. Thus, the non-zero one-step state transition probabilities for any $0\!<\!i_0\!<\!c, 0\!<\!j_0\!<\!c,{\rm and~} 0\!<\!k_0\!<\!h$ are given as follows:
\vspace{-0.05in}
\begin{equation}
\vspace{-0.05in}
\left\{
\begin{array}{ll}
P(0,0,k_0|0,0,k_0)=qu+(1-u) \\
P(1,0,k_0|0,0,k_0)=u(1-p)(1-q) \\
P(0,1,k_0|0,0,k_0)=up(1-q) \\
P(i_0,j_0+1,k_0|i_0,j_0,k_0)=1 \\
P(i_0,1,k_0|i_0,0,k_0)=p \\
P(i_0+1,0,k_0|i_0,0,k_0)=1-p\\
P(1,0,k_0+1|c,0,k_0)=1-p\\
P(0,1,k_0+1|c,0,k_0)=p\\
P(0,0,0|c,0,h)=1-s\\
P(0,0,1|c,0,h)=s\\
P(0,0,0|0,0,0)=1-s\\
P(0,0,1|0,0,0)=s
\end{array}
\right.
\vspace{-0.02in}
\end{equation}

Let $P_{(i,j,k)}\!\!=\!\!\lim_{t\to\infty}P(N_t(t)\!\!=\!\!i,N_c(t)\!\!=\!\!j,N_f(t)\!\!=\!\!k),i\!\in\![0,c],j\!\in\![0,c],k\!\in\![0,h]$ be the steady-state probability of the Markov chain. We first study a simple case where no PU exists in the CR network. Then, we consider the scenario where SUs coexist with PUs.

\subsubsection{No PU Exists in a Network}
\label{sssc:nopu}
In this case, since the probability that a PU packet arrives in a time slot is equal to zero (i.e., $p\!=\!0$), all channels are always available for SUs (i.e., $u$=1) and a SU does not need to perform spectrum handoffs during a data transmission. Thus, a SU cannot be in the \textit{Collided} state. In addition, a SU can only be in the \textit{Backlogged} state when it initiates a new transmission (i.e., the \textit{Backlogged} states are reduced to $(N_t(t)\!\!=\!\!0,N_c(t)\!\!=\!\!0,N_f(t)\!\!=\!\!1)$. Thus, the steady-state probabilities of the \textit{Transmitting} and \textit{Idle} state can be represented in terms of the steady-state probability of the \textit{Backlogged} state $P_{(0,0,1)}$. Hence, from Fig. \ref{fig:markov},
\begin{equation}\label{eq:eq1}
P_{(i,0,k)}=(1-q)P_{(0,0,1)}, {\rm ~for~} 1\leq i\leq c, 1\leq k\leq h,
\end{equation}
\begin{equation}\label{eq:eq2}
P_{(0,0,0)}=\frac{(1-s)(1-q)}{s}P_{(0,0,1)}.
\end{equation}
Since $\sum_i\sum_j\sum_kP_{(i,j,k)}\!\!=\!\!1$, we can calculate the steady-state probability of every state in the Markov chain. Note that the probability of a collision among SUs, $q$, depends on the channel selection scheme. The derivation of $q$ is given in Section \ref{sc:selection}.

\subsubsection{SUs Coexist with PUs in a Network}
\label{sssc:pu}
If the probability that a PU packet arrives in a time slot is not equal to zero (i.e., $p\!\neq\!0$), collisions between SUs and PUs may occur when a SU transmits a frame. Thus, the steady-state probabilities of the \textit{Collided} states are not zero. Similar to the no-PU case, we represent the steady-state probabilities in terms of $P_{(0,0,1)}$. First of all, for the first tier in Fig. \ref{fig:markov}, we can obtain the steady-state probabilities of all the \textit{Transmitting} states in terms of $P_{(0,0,1)}$, that is,
\begin{equation}\label{eq:tier1trans}
P_{(i,0,1)}=u(1-q)(1-p)^{i}P_{(0,0,1)}, {\rm ~for~} 1\leq i\leq c.
\end{equation}
Then, for the \textit{Collided} states with $i=0$,
\begin{equation}\label{eq:tier1col1}
P_{(0,j,1)}=up(1-q)P_{(0,0,1)}, {\rm ~for~} 1\leq j\leq c.
\end{equation}
For the \textit{Collided} states with $i>0$,
\begin{equation}\label{eq:tier1col2}
P_{(i,j,1)}\!=\!u(1\!-\!q)p(1\!-\!p)^{i}P_{(0,0,1)}, {\rm ~for~} 1\!\leq\! i\!\leq\! c\!-\!1, 1\!\leq\! j\!\leq\! c.
\end{equation}
For the $k$-th $(k>1)$ tier, we first derive $P_{(1,0,k)}$ and $P_{(0,1,k)}$:
\begin{equation}\label{eq:tierktran1}
P_{(1,0,k)}=(1-p)P_{(c,0,k-1)}+u(1-p)(1-q)P_{(0,0,k)},
\end{equation}
\begin{equation}\label{eq:tierkcol1}
P_{(0,1,k)}=pP_{(c,0,k-1)}+up(1-q)P_{(0,0,k)}.
\end{equation}
Then, the steady-state probabilities of the \textit{Transmitting} states when $i>1$ can be represented as
\begin{equation}\label{eq:tierktrans}
P_{(i,0,k)}=(1-p)^{i-1}P_{(1,0,k)}, {\rm ~for~} 1<i\leq c.
\end{equation}
Similar to the derivation method for the first tier, for the \textit{Collided} states with $i=0$,
\begin{equation}\label{eq:tierkcol2}
P_{(0,j,k)}=P_{(0,1,k)}, {\rm ~for~} 1\leq j\leq c.
\end{equation}
For the \textit{Collided} states with $i>0$,
\begin{equation}\label{eq:tierkcol3}
P_{(i,j,k)}\!=\!p(1\!-\!p)^{i\!-\!1}P_{(1,0,k)}, {\rm ~for~} 1\!\leq\! i\!\leq\! c-1, 1\!\leq\! j\!\leq\! c.
\end{equation}
Then, for the $Backlogged$ state in the $k$-th tier,
\vspace{-0.05in}
\begin{equation}\label{eq:tierkback}
\sum_{i=0}^{c-1}P_{(i,c-i,k)}=u(1-q)P_{(0,0,k)}.
\vspace{-0.05in}
\end{equation}
Combining (\ref{eq:tierktran1}) through (\ref{eq:tierkback}), we obtain the following equations using basic mathematical manipulations:
\begin{equation}\label{eq:tierktrans1}
P_{(1,0,k)}=\frac{1}{(1-p)^{c-1}}P_{(c,0,k-1)},
\end{equation}
\begin{equation}\label{eq:tierkcol11}
P_{(0,1,k)}=\frac{p}{(1-p)^c}P_{(c,0,k-1)},
\end{equation}
\begin{equation}\label{eq:tierkback1}
P_{(0,0,k)}=\frac{1-(1-p)^c}{u(1-q)(1-p)^c}P_{(c,0,k-1)}.
\end{equation}
Then, from (\ref{eq:tierktrans}),
\vspace{-0.05in}
\begin{equation}\label{eq:recur1}
P_{(c,0,k-1)}=(1-p)^{c-1}P_{(1,0,k-1)}.
\vspace{-0.05in}
\end{equation}
Combining (\ref{eq:tierktrans1}) and (\ref{eq:recur1}), we find the following relationship:
\begin{equation}\label{eq:recur2}
P_{(c,0,k)}=P_{(c,0,k-1)}.
\end{equation}
Thus,
\vspace{-0.05in}
\begin{equation}
\vspace{-0.05in}
\label{eq:recur3}
P_{(c,0,k)}=u(1-q)(1-p)^cP_{(0,0,1)}.
\end{equation}
(\ref{eq:recur3}) indicates the steady-state probabilities of the states in the $k$-th tier are independent of $k$. Now, we have all the steady-state probabilities of the states in all tiers except the state $(0,0,0)$. At last, for the $Idle$ state,
\begin{equation}\label{eq:idle}
P_{(0,0,0)}=\frac{1-s}{s}u(1-q)(1-p)^cP_{(0,0,1)}.
\end{equation}
Similarly, since $\sum_i\sum_j\sum_kP_{(i,j,k)}=1$, we can get the steady-state probability of every state in the Markov chain. If we denote $\Theta$ as the normalized throughput of SU transmissions, $\Theta$ is the summation of the steady-state probabilities of all the $Transmitting$ states in our proposed Markov model. That is,
\vspace{-0.1in}
\begin{equation}\label{eq:throu}
\vspace{-0.05in}
\Theta=\sum_{k=1}^h\sum_{i=1}^cP_{(i,0,k)}. 
\end{equation}

\subsection{The Probability that at Least One Channel is Idle}
\label{ssc:puchannel}
In the above derivations, $u$ and $q$ are unknown. In this subsection, we calculate the probability that at least one channel is idle, $u$. In this paper, we only consider homogeneous PU traffic on each channel. Without loss of generality, we associate each PU with one channel and model the activity of each PU as an ON/OFF process \cite{SWanginfocom10}\cite{Su08}\cite{Zhang06}. SUs can only exploit the channels when the channels are idle (i.e., in the OFF period). We assume that the buffer in each PU can store at most one packet at a time. Once a packet is stored at a buffer, it remains there until it is successfully transmitted. Thus, we assume that the OFF period of a channel follows the geometric distribution, where the probability mass function (pmf) is given by 
\vspace{-0.02in}
\begin{equation}
\Pr(N_{OFF}=n)=p(1-p)^{n},
\label{eq:interarrival}
\end{equation}
where $N_{OFF}$ is the number of time slots of an OFF period.

\begin{figure}[htb!]
\vspace{-0.12in}
	\centering
		\includegraphics[width=0.3\textwidth]{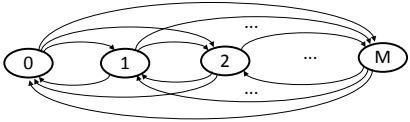}
		\vspace{-0.05in}
	\caption{The transition diagram of the number of channels used by PUs in one time slot.}
	\label{fig:markov2}
	\vspace{-0.12in}
\end{figure}

Let $\Omega(t)$ be the number of channels used by PUs at time slot $t$. The process $\{\Omega(t),t=0,1,2,\cdots\}$ forms a Markov chain whose state transition diagram is given in Fig. \ref{fig:markov2}, in which the self loops are omitted. To characterize the behavior of the PU channels, we define $\mathcal{D}_\alpha^l$ as the event that $l$ PUs finish their transmissions given that there are $\alpha$ PUs in the network in a time slot. We also define $\mathcal{A}_\gamma^m$ as the event that $m$ PUs start new transmissions given that there are $\gamma$ idle PUs in a time slot. Thus, the probabilities of events $\mathcal{D}_\alpha^l$ and $\mathcal{A}_\gamma^m$ are:
\begin{equation}\label{eq:eventa}
\Pr(\mathcal{D}_\alpha^l)=\binom{\alpha}{l}v^l(1-v)^{\alpha-l},
\end{equation}
\begin{equation}\label{eq:eventb}
\Pr(\mathcal{A}_\gamma^m)=\binom{\gamma}{m}p^m(1-p)^{\gamma-m},
\end{equation}
where $v$ is the probability that a PU finishes its transmission in a slot. If the average length of a PU packet is denoted as $\bar{L}$, then $v\!=\!1/\bar{L}$. Therefore, the state transition probability from state $\{\Omega(t)\!=\!a\}$ to state $\{\Omega(t\!+\!1)\!=\!b\}$ can be written as
\begin{equation}
p_{ab} = 
\left\{
\begin{array}{ll}
\sum_{l=0}^a\Pr(\mathcal{D}_a^l)\Pr(\mathcal{A}_{M-a+l}^{b-a+l}), &{\rm for~} b\geq a \\
\sum_{l=a-b}^a\Pr(\mathcal{D}_a^l)\Pr(\mathcal{A}_{M-a+l}^{b-a+l}), &{\rm for~} b< a.
\end{array}
\right.
\label{eq:trans}
\end{equation}
Therefore, we can obtain the steady-state probabilities of the number of busy channels in the band in a time slot, denoted as $\textbf{g}=[g_0 ~~ g_1~~ g_2~\cdots~ g_M]^T$, where $g_i$ denotes the steady-state probability that there are $i$ busy channels in a time slot. Hence, $u=\sum_{i=0}^{M-1}g_i$.
\vspace{-0.1in}

\section{The Impact of Different Channel Selection Schemes}
\label{sc:selection}
In this section, we investigate the impact of different channel selection schemes on the performance of the spectrum handoff process in a multi-SU scenario by deriving the probability of a collision among SUs, $q$.
\subsection{Random Channel Selection}
\label{ssc:random}
For the random channel selection scheme, a SU selects an available channel for access on a random basis. Thus, a collision among SUs happens if more than one SU selects the same channel. To make the analysis tractable, we assume that the SU traffic is saturated (i.e., after finishing transmitting a packet, a SU always has another packet in the buffer to send). Thus, let $\mathcal{B}(t)$, $\mathcal{T}(t)$, and $\mathcal{C}(t)$ be the number of SUs in the \textit{Backlogged}, \textit{Transmitting}, and $Collided$ state at time slot $t$, respectively.  Therefore, $\mathcal{B}(t)\!+\!\mathcal{T}(t)\!+\!\mathcal{C}(t)\!=\!N$. The process $\{\mathcal{B}(t),\mathcal{T}(t),\mathcal{C}(t),t=1,2,\cdots\}$ forms a Markov chain, namely the system chain. Thus, we denote the state transition probability of the system chain from $(n_1,n_2,n_3)$ to $(n_1',n_2',n_3')$ as $K_{((n_1,n_2,n_3),(n_1',n_2',n_3'))}$. Let $X_w(n_2)$ be the probability that $w$ number of SUs in the $Transmitting$ state successfully finish their transmissions at time slot $t$ given that there are $n_2$ SUs in the $Transmitting$ state. Then,
\begin{equation}
X_w(n_2)=\binom{n_2}{w}\sigma^w(1-\sigma)^{n_2-w},
\label{eq:trans2back}
\end{equation}
where $\sigma$ is the probability that a SU finishes a packet transmission in a slot. Let $Y_r(n_2,w)$ be the probability that $r$ SUs in the $Transmitting$ state collide with PU packets in the next time slot given that $n_2$ SUs are in the $Transmitting$ state and $w$ SUs out of $n_2$ SUs finish their transmissions. Thus,
\begin{equation}
Y_r(n_2,w)=\binom{n_2-w}{r}p^r(1-p)^{n_2-w-r}.
\label{eq:trans2col}
\end{equation}
Let $Z_e(n)$ be the probability that $e$ of $n_3$ users transmit the last time slot of a frame in the current slot given there are $n_3$ SUs in the $Collided$ state. Then,
\begin{equation}
Z_e(n_3)=\binom{n_3}{e}p_f^e(1-p_f)^{n_3-e},
\label{eq:col2back}
\end{equation}
where $p_f$ is the probability that the current time slot is the end of a frame. Since the frame length is $c$ time slots, $p_f\!=\!\frac{1}{c}$. Let $T_d(n_1,\theta)$ be the probability that $d$ SUs successfully access the channels given that there are $n_1$ SUs in the $Backlogged$ state and $\theta$ available channels in the band. Then,
\begin{equation}\label{eq:bak2trans1}
T_d(n_1,\theta)=\frac{S_{d}(n_1,\theta)}{\binom{\theta+n_1-1}{n_1}},
\end{equation}
where $S_d(n_1,\theta)$ is the number of possibilities that $d$ of $n_1$ SUs select a channel that is only selected by one SU given that there are $\theta$ channels available. The denominator in (\ref{eq:bak2trans1}) is the total number of possibilities that $n_1$ SUs select $\theta$ available channels. $S_d(n_1,\theta)$ can be calculated using the following iterative equation:
\begin{equation}\label{eq:bak2trans}
\begin{split}
S_d(n_1,\theta)=& U_d(n_1,\theta)\!-\!U_{d+1}(n_1,\theta)-\\
&\sum_{i=1}^{n_1-d}\left[\binom{d+i}{d}\!-\!\binom{d+i}{d+1}\right]S_{d+i}(n_1,\theta),
\end{split}
\end{equation}
where $U_d(n_1,\theta)\!=\!\binom{n_1}{d}\binom{\theta+n_1-2d-1}{\theta-d}$. The proof of (\ref{eq:bak2trans}) is given in the Appendix. Since $n_2\!=\!N\!-\!n_1\!-\!n_3$, we can remove $n_2$ from the state space and reduce the state space from three dimensions to two dimensions. Thus, the system chain becomes a two-dimensional Markov chain $\{\mathcal{B}(t),\mathcal{C}(t)\}$. The state transition probability is 
\begin{equation}\label{eq:statetrans}
\begin{split}
K_{((n_1,n_3),(n_1',n_3'))}=&\sum_{\theta=0}^M\sum_{e=0}^{n_3}\sum_{w=0}^{N\!-\!n_1\!-\!n_3}T_{n_1\!-\!n_1'\!+\!w\!+\!e}(n_1,\theta)\\
&Y_{n_3'+e-n_3}(N-n_1-n_3,w)\\
&X_w(N-n_1-n_3)Z_e(n_3)\Pr(\theta),
\end{split}
\end{equation}
where $\Pr(\theta)$ is the steady-state probability that there are $\theta$ channels available in the band, which can be obtained in Section \ref{ssc:puchannel}.

We further reduce the two dimensional system chain $\{\mathcal{B}(t),\mathcal{C}(t)\}$ with the state transition probability matrix $K_{((n_1,n_3),(n_1',n_3'))}$ to a one dimensional Markov chain with the state transition probability matrix $H_{(m,m')}\!=\!K_{((n_1,n_3),(n_1',n_3'))}$, where
\vspace{-0.05in}
\begin{equation}\label{eq:reducetrans}
\left\{
\begin{array}{lll}
m &=&\frac{(2N-n_1+3)n_1}{2}+n_3 \\
m'&=&\frac{(2N-n_1'+3)n_1'}{2}+n_3'.
\end{array}
\right.
\end{equation}
Let $\pi_{m}$ be the steady-state probability for state $m, 0\leq m\leq\frac{(N+1)(N+2)}{2},$ of the one-dimensional Markov chain with the state transition probability matrix $H_{(m,m')}$. By solving the equilibrium equation $\pi_m'\!=\!\sum_{m=0}^{\frac{(N+1)(N+2)}{2}}\pi_mH_{(m,m')}$ with the condition $\sum_{m=0}^{\frac{(N+1)(N+2)}{2}}\pi_m\!=\!1$, we can obtain the steady-state probability $\pi_m$. We denote the steady-state probability that there are $k$ SUs in the $Backlogged$ state as $\rho_k$. $\rho_k$ can be calculated by adding all the $\pi_m$ in which $m$ should be:
\begin{equation}\label{eq:rho}
m=\frac{(2N-k+3)k}{2}+j, {\rm~~} \forall j\in[0,N-k].
\end{equation}
Thus,
\vspace{-0.05in}
\begin{equation}\label{eq:steadyback}
\rho_k=\sum_{m=\frac{(2N\!-\!k\!+\!3)k}{2}}^{\frac{(2N\!-\!k\!+\!3)k}{2}\!+\!N\!-\!k}\pi_m.
\vspace{-0.05in}
\end{equation}
Thus, the probability that a collision occurs among SUs when they randomly select a channel for each SU is obtained by
\vspace{-0.05in}
\begin{equation}\label{eq:q1}
q=\sum_{\theta=1}^M\sum_{k=1}^N\frac{k-1}{\theta+k-2}\rho_k\Pr(\theta).
\end{equation}

\vspace{-0.05in}
\subsection{Greedy Channel Selection}
\label{ssc:greedy}
\begin{figure}[htb!]
\vspace{-0.1in}
	\centering
		\includegraphics[width=0.35\textwidth]{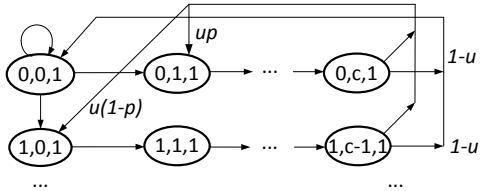}
		\vspace{-0.1in}
	\caption{The modified Markov model based on the greedy channel selection scheme.}
	\label{fig:greedyselection}
	\vspace{-0.18in}
\end{figure}

For the greedy channel selection scheme, a SU always selects the channel which leads to the minimum service time \cite{LCWang09}. If more than one SU pair perform spectrum handoffs at the same time, this channel selection method will cause definite collisions among SUs. Thus, the probability that a collision occurs among SUs is given by:
\vspace{-0.05in}
\begin{equation}\label{eq:q2}
q=
\left\{
\begin{array}{ll}
0 &{\rm for~} N = 1 \\
1 & {\rm for~} N>1.
\end{array}
\right.
\vspace{-0.05in}
\end{equation}
Note that in this channel selection scheme, both the SU transmitter and receiver do not need to exchange information on the selected channel. Thus, the transition probability from the $Collided$ states to the corresponding $Backlogged$ state is $1\!-\!u$ instead of one. In addition, under homogeneous PU traffic, the greedy channel selection scheme is equivalent to the random channel selection scheme. A part of the modified state transition diagram for the first tier is shown in Fig. \ref{fig:greedyselection}. The derivation of the steady-state probabilities of this modified model can be carried out in the way as in Section \ref{ssc:prob}.

\subsection{Pseudo-Random Selecting Sequence based Channel Selection}
\label{ssc:sequence}
A channel selection scheme is proposed based on a pseudo-random selecting sequence \cite{SongGC10}. When multiple SUs perform spectrum handoffs at the same time, a pseudo-random selecting sequence for each SU is generated locally. SUs need to perform spectrum handoffs following the same selecting sequence to select channels to avoid collisions. Thus, for this channel selection scheme, the probability of a collision among SUs is always zero (i.e., $q\!=\!0$).

\vspace{-0.05in}
\subsection{Results Validation}
\label{ssc:validation}
\begin{figure}[htb!]
\vspace{-0.16in}
	\centering
		\includegraphics[width=0.4\textwidth]{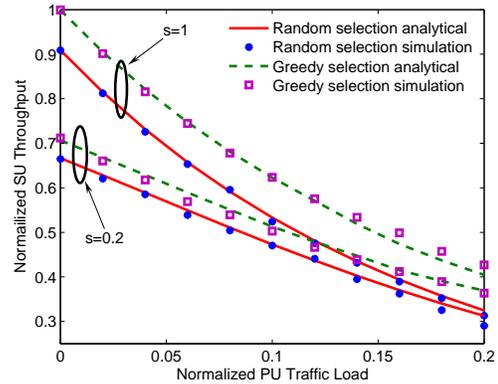}
		\vspace{-0.15in}
	\caption{Analytical and simulation results of the normalized SU throughput in a two-SU scenario.}
	\label{fig:twousers}
	\vspace{-0.12in}
\end{figure}

In this subsection, we validate the numerical results obtained from our proposed Markov model using simulation. Note that when the number of SUs in the network is larger than two, the throughput using the greedy channel selection scheme for spectrum handoff is always zero because $q\!=\!\!1$. Thus, we first validate our numerical results in a two-SU scenario, where the number of PU channels, $M\!=\!10$. The number of frames in a SU packet, $h\!=\!1$, and the number of slots in a frame, $c\!=\!10$. We assume that the SU packets are of fixed length. Thus, $\sigma\!=\!\frac{1}{ch}$. Fig. \ref{fig:twousers} depicts the analytical and simulation results of the normalized SU throughput using the random channel selection scheme and the greedy channel selection scheme. In the simulation, only data frames are considered for the calculation of the throughput. Signaling packets (RTS/CTS) are not. It can be seen that the simulation results match extremely well with the numerical results in both schemes with the maximum difference only $3.84\%$ for the random selection and $4.09\%$ for the greedy selection. It is also shown that, under the same SU traffic load, the greedy channel selection scheme always outperforms the random channel selection scheme in terms of higher SU throughput.

\begin{figure}[htb!]
\vspace{-0.17in}
	\centering
		\includegraphics[width=0.4\textwidth]{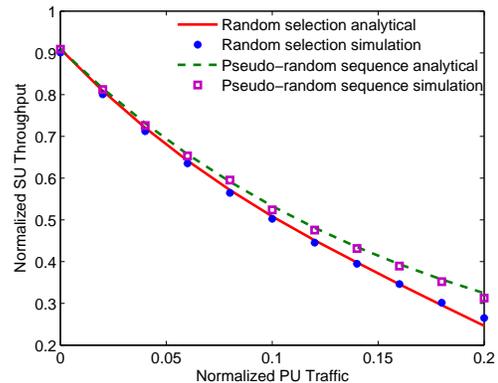}
		\vspace{-0.14in}
	\caption{Analytical and simulation results of the normalized SU throughput under different channel selection schemes in a twenty-SU scenario.}
	\label{fig:channelselection}
			\vspace{-0.15in}
\end{figure}

Then, we consider a network with 10 SU pairs in the network. We fixed the SU traffic at $s\!=\!1$. The rest of the parameters are the same as in the two-SU scenario. Fig. \ref{fig:channelselection} shows that, under different channel selection schemes, the analytical and simulation results match well with the maximum difference only $6.14\%$ for the random selection and $1.2\%$ for the pseudo-random sequence selection. Fig. \ref{fig:channelselection} also indicates that the pseudo-random sequence selecion outperfoms the random selection, especially when PU traffic is high.

\section{The Impact of Spectrum Sensing Delay}
\label{sc:sensing}
In this section, we investigate the impact of the spectrum sensing delay on the performance of a spectrum handoff process. The spectrum sensing delay considered in this paper is defined as the duration from the moment that a collision between a SU and PU happens to the moment that the SU detects the collision (i.e., the overlapping time between a SU and PU transmission). Let $T_s$ be the spectrum sensing delay. Therefore, a SU does not need to wait till the last time slot of a frame to realize the collision, as shown in Fig. \ref{fig:handoff}. It only needs to wait for $T_s$ to realize that a collision with a PU packet occurs and stops the current transmission immediately. In a recent work \cite{CWWangGC10}, the spectrum sensing time is considered as a part of the spectrum handoff delay. However, the definition of the spectrum sensing time in \cite{CWWangGC10} is different from the definition considered in this paper. In \cite{CWWangGC10}, the spectrum sensing time only refers to the duration that a SU finds an available channel for transmission after a collision occurs. Thus, the spectrum sensing time can be as low as zero in \cite{CWWangGC10}. In addition, the overlapping time of a SU and PU collision is neglected in \cite{CWWangGC10}. However, the spectrum sensing delay considered in this paper is not negligible. 

\begin{figure}[htb!]
\vspace{-0.12in}
	\centering
		\includegraphics[width=0.45\textwidth]{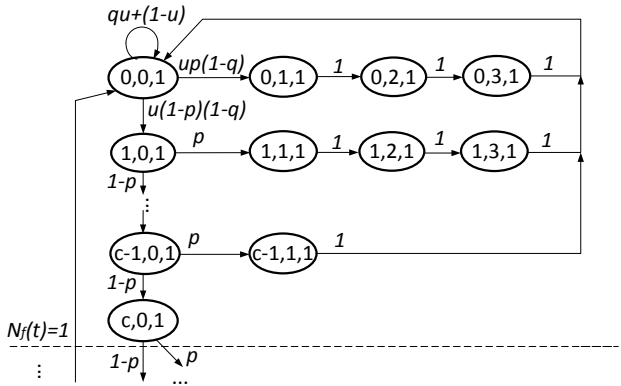}
		\vspace{-0.07in}
	\caption{The modified Markov model based on the spectrum sensing delay when $T_s$ equals 3 time slots.}
	\label{fig:sensingdelay}
	\vspace{-0.12in}
\end{figure}

The spectrum sensing delay, $T_s$, can be easily implemented in our proposed three dimensional Markov model with minor modifications. Fig. \ref{fig:sensingdelay} shows the first tier of the modified three dimensional discrete-time Markov chain when $T_s$ equals 3 time slots. It is shown that, for a fixed $N_t(t)$, the maximum number of $Collided$ states is $T_s$. The modified model of other tiers is similar to the first tier as shown in Fig. \ref{fig:sensingdelay}.

Compared with the original Markov model shown in Fig. \ref{fig:markov}, the derivation of the steady-state probabilities of the Markov model implemented with the spectrum sensing delay is exactly the same. The only difference is that the total number of the \textit{Collided} states in the modified Markov model is reduced from $[c(c\!+\!1)/2]h$ in the original Markov model to $[T_s(c\!-\!T_s\!+\!1)\!+\!T_s(T_s\!-\!1)/2]h$.

\begin{figure}[htb!]
\vspace{-0.15in}
	\centering
		\includegraphics[width=0.4\textwidth]{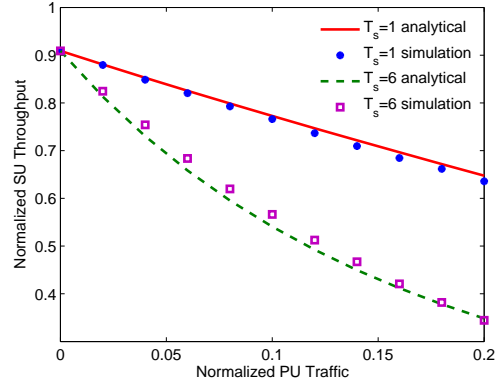}
		\vspace{-0.07in}
	\caption{Analytical and simulation results of the normalized SU throughput under different spectrum sensing delay.}
	\label{fig:sensing}
	\vspace{-0.14in}
\end{figure}

Fig. \ref{fig:sensing} shows the impact of the spectrum sensing delay on the SU throughput performance. We consider a two-SU scenario with different spectrum sensing delay using the random channel selection scheme. It is shown that the numerical results and analytical results match well with the maximum difference $1.83\%$ for $T_s\!=\!1$ and $4.56\%$ for $T_s\!=\!6$. It reveals that our proposed model can accurately predict the SU throughput. 

\section{Performance Evaluation}
\label{sc:evaluation}
In this section, we use our proposed Markov model to evaluate the performance of SU transmissions in spectrum handoff scenarios under various system parameters.
\subsection{Collision Probability between SUs and PUs}
\label{ssc:colprob}
\begin{figure}[htb!]
\vspace{-0.12in}
	\centering
		\includegraphics[width=0.4\textwidth]{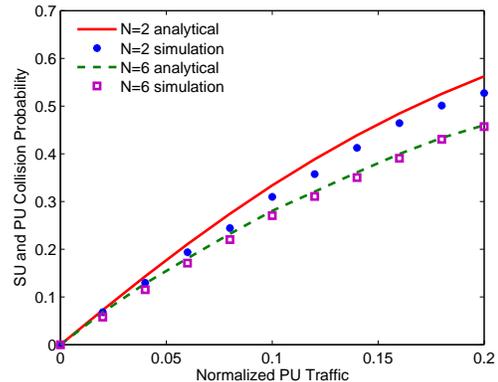}
		\vspace{-0.07in}
	\caption{Analytical and simulation results of the collision probability between SUs and PUs.}
	\label{fig:colprob}
	\vspace{-0.12in}
\end{figure}

Based on the proposed Markov model, the collision probability between SUs and PUs is the summation of all the steady-state probabilities of the $Collided$ states. That is, $\Pr[collision]=\sum_{k=1}^h\sum_{i=0}^{c-1}\sum_{j=1}^{c-i}P_{(i,j,k)}$. Fig. \ref{fig:colprob} shows the analytical and simulation results of the collision probability between SUs and PUs using the random channel selection scheme. The analytical results fit simulation results well with the maximum difference $6.26\%$ for $N\!=\!2$ and $3.41\%$ for $N\!=\!6$, respectively. It is shown that the collision probability between SUs and PUs decreases as the number of SUs increases. This is because that the number of collisions among SUs increases as the number of SUs during a spectrum handoff increases. Therefore, the probability for a SU being in the $Backlogged$ states increases. Thus, the collision probability between SUs and PUs drops.

\subsection{Average Spectrum Handoff Delay}
\label{ssc:handoffdelay}
\begin{figure}[htb!]
\vspace{-0.15in}
	\centering
		\includegraphics[width=0.4\textwidth]{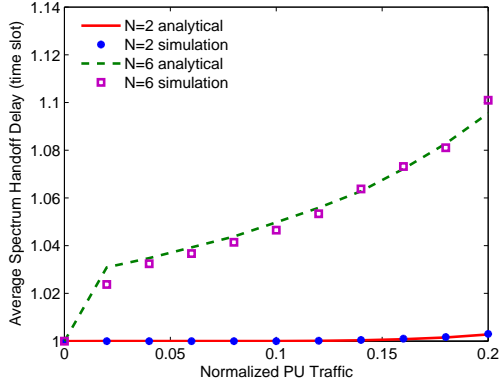}
		\vspace{-0.1in}
	\caption{Analytical and simulation results of the average spectrum handoff delay.}
	\label{fig:handoffdelay}
	\vspace{-0.2in}
\end{figure}

We denote $D_s$ as the average spectrum handoff delay. Since the spectrum handoff delay is equivalent to the dwelling time on the $Backlogged$ state, we obtain
\begin{equation}\label{eq:handoffdelay}
D_s = \sum_{k=1}^\infty kp_d^{k-1}(1-p_d),
\end{equation}
where $p_d\!=\!qu\!+\!(1\!-\!u)$. Fig. \ref{fig:handoffdelay} shows the analytical and simulation results of the average spectrum handoff delay using the random channel selection scheme. It is shown that as the number of SUs increases, the average spectrum handoff delay increases drastically.
\vspace{-0.08in}

\section{Conclusion}
\label{sc:conclusion}
In this paper, a novel three dimensional discrete-time Markov chain is proposed to analyze the performance of SUs in the spectrum handoff scenario in a CR ad hoc network under homogeneous PU traffic. We performed extensive simulations in different network scenarios to validate our proposed model. The analysis shows that our proposed Markov model is very flexible and can be applied to various practical network scenarios. Thus, our analysis provides insights into the spectrum handoff process for CR networks. This allows us to obtain the throughput and other performance metrics for various design requirements. Currently, no existing analysis has considered the comprehensive aspects of spectrum handoff as what we considered in this paper. Finally, although we focus on the spectrum handoff scenario in CR networks, the modeling techniques developed in the paper are quite general and are applicable to other multi-channel scenarios with multiple interacting users.

\vspace{-0.05in}

\appendix
We now give the proof of (\ref{eq:bak2trans}). $S_d(n_1,\theta)$ is the number of possibilities that $d$ channels are selected by only one SU for each channel given that there are $n_1$ SUs and $\theta$ available channels. Let $\Phi_d(n_1,\theta)$ be the number of possibilities that at least $d$ channels are selected by one SU for each channel given that there are $n_1$ SUs and $\theta$ available channels. Thus, 
\vspace{-0.02in}
\begin{equation}\label{eq:appen1}
S_d(n_1,\theta)=\Phi_d(n_1,\theta)-\Phi_{d+1}(n_1,\theta).
\end{equation}

Then, we calculate $\Phi_d(n_1,\theta)$. We first select $d$ channels out of $n_1$ with one SU on each channel. The total number of possibilities is $\binom{n_1}{d}$. Then, let the remaining $n_1-d$ SUs select the remaining $\theta-d$ channels. The total number of possibilities is $\binom{\theta+n_1-2d-1}{\theta-d}$. Thus, we denote $U_d(n_1,\theta)=\binom{n_1}{d}\binom{\theta+n_1-2d-1}{\theta-d}$. Compare $\Phi_d(n_1,\theta)$ with $U_d(n_1,\theta)$, there are many repeated counts that need to be removed. We denote the number of repeated counts as $\Gamma_d(n_1,\theta)$.

Note that for the $d+i,i\!>\!0,$ channels that are selected by only one SU, the number of repeated counts is $\left[\binom{d+i}{d}\!-\!1\right]S_{d+i}(n_1,\theta)$. Thus, the total number of repeated counts is
\vspace{-0.05in}
\begin{equation}\label{eq:appen2}
\vspace{-0.05in}
\Gamma_d(n_1,\theta)=\sum_{i=1}^{n_1-d}\left[\binom{d+i}{d}-1\right]S_{d+i}(n_1,\theta).
\end{equation}
Thus,
\begin{equation}\label{eq:appen3}
\Phi_d(n_1,\theta)=U_d(n_1,\theta)-\Gamma_d(n_1,\theta).
\end{equation} 
Compare (\ref{eq:appen1}) through (\ref{eq:appen3}), (\ref{eq:bak2trans}) is obtained.

\end{document}